\documentclass[%
reprint,
superscriptaddress,
 amsmath,amssymb,
 aps,
pra,
]{revtex4-1}

\usepackage{graphicx}
\usepackage{dcolumn}
\usepackage{bm}
\usepackage{color} 
\newcommand{\beq}{\begin{equation}}
\newcommand{\eeq}{\end{equation}}
\newcommand{\beqa}{\begin{eqnarray}}
\newcommand{\eeqa}{\end{eqnarray}}

\begin{document}
\preprint{AIP/123-QED}

\title
{{Electronic band gap and transport in Fibonacci quasi-periodic graphene superlattice}}

\author{Pei-Liang Zhao}
\affiliation{Department of Physics, Shanghai University,
200444 Shanghai, P. R. China}

\author{Xi Chen}
\email[Corresponding author. Email:~]{xchen@shu.edu.cn}
\affiliation{Department of Physics, Shanghai University,
200444 Shanghai, P. R. China}

\affiliation{Departamento de Qu\'{\i}mica F\'{\i}sica, UPV-EHU, Apdo 644, E-48080 Bilbao, Spain}

\date{\today}

\begin{abstract}
We investigate electronic band gap and transport in Fibonacci quasi-periodic graphene superlattice.
It is found that such structure can
possess a zero-$\bar{k}$ gap which exists in all Fibonacci
sequences. Different from Bragg gap, zero-$\bar{k}$ gap associated with Dirac point is less sensitive to the incidence angle and lattice constants. The defect mode appeared inside
the zero-$\bar{k}$ gap has a great effect on transmission, conductance and shot noise,
which can be applicable to control the electron transport.
\end{abstract}


\maketitle


Graphene, a monolayer of carbon atoms tightly packed into a honeycomb lattice, has attracted great interest in graphene-based nanoelectronic and optoelectronic devices
\cite{Castro},
since it was fabricated by Novoselov and Geim \textit{et al.} in 2004 \cite{Novoselov}. In graphene, the unique band structure with the valance and conduction bands touching at Dirac point (DP)
leads to the fact that electrons around the Fermi level can be described as the massless relativistic Dirac
electrons, resulting in the linear energy dispersion relation. As a consequence, there are a great number of
electronic properties, such as the half-integer
quantum Hall effect \cite{Novoselov-GMJK,Zhang-TS,Gusynin-S}, the
minimum conductivity \cite{Novoselov-GMJK}, and Klein tunneling \cite{Katsnelson-NG}.
In particular, Klein tunneling and perfect transmission are crucial for electron transport in
various graphene heterostructures \cite{Young}, i.e. single barrier \cite{Chen-APL} and n-p-n junctions \cite{Cheianov}.

Motivated by the experimental realization of graphene superlattice (GSL) \cite{Meye,Marchini,Vazquez},
electronic bandgap structures and transport properties
in GSLs with electrostatic potential and magnetic barrier have
been extensively investigated \cite{Bai,Barbier2009,Brey,Park2008,Wang2010,YuXian,Abedpour,Bliokh,Mukhopadhyay,Sena},
since the conventional semiconductor superlattices are successful
in controlling the electronic structures and the extension to graphene may give rise to
different features and applications. For instance, DP appears
in the GSL  \cite{Barbier2009,Brey}, and it is exactly located at
the energy with the zero-$\bar{k}$ gap \cite{Wang2010}. Interestingly,
the zero-$\bar{k}$ gap associated with DP is insensitive to the lattice parameter changes in contrast with the behavior exhibited by Bragg gaps \cite{Wang2010}.
This gap is analogous to photonic zero-$\bar{n}$ gap in the photonic crystals containing negative-index and
positive-index materials \cite{Bliokh}, and originates from a zero total phase \cite{Li-PRL}.
Accordingly, the zero-$\bar{k}$ gap is robust against the lattice constants, structural disorder \cite{Wang2010},
and external magnetic field \cite{YuXian}, and thus is better to control the electron transport in GSL.

%
%
\begin{figure}[]
\begin{center}
\scalebox{0.50}[0.45]{\includegraphics{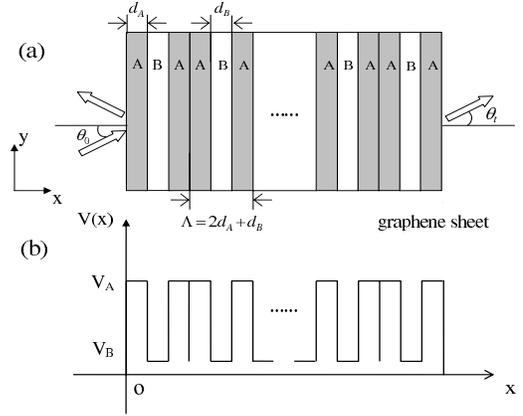}}
\caption{(Color online) (a) Example of quasi-periodic monolayer GSL, corresponding to Fibonacci sequence $S_3$. (b) The schematic profiles of the potentials $V_{A}$ and $V_B$.}
\label{Fig.1}
\end{center}
\end{figure}
%
In this Letter, we will investigate electronic band gap and transport in Fibonacci quasi-periodic GSLs in the fashion analogous to photonic crystal with metamaterials
\cite{Li-PRL,Alfonoso,Zhang}. As we know, the quasi-periodic GSL
is classified as intermediate between ordered and disordered systems \cite{Abedpour,Bliokh}, which has significant and common
features like fractal spectrum and self-similar behavior \cite{Mukhopadhyay,Sena}. However, what we concentrate on here is the
electronic band gap and DP in such quasi-periodic system. We find that
zero-$\bar{k}$ gap happens in all Fibonacci sequences, 
which results in the robust transmission properties, conductance and shot noise at the DP.

Consider quasi-periodic monolayer GSLs
with the structure in each cell following the Fibonacci sequence, $S_{j}$,
by a recurrent relation $S_{j+1}=\{S_{j},S_{j-1}\}$, with
$S_{0}=\{B\}$ and $S_{1}=\{A\}$ with $j$ is the generation number of the Fibonacci unit cell, the first few sequences are $S_{2}=\{AB\}$,
$S_{3}=\{ABA\}$, $S_{4}=\{ABAAB\}$ and so on.  Elements $A$ and $B$ are
considered as the alternating barriers $V_A$ and wells $V_B$ with the width $d_{A}$ and $d_{B}$,
respectively. 
As an example, the third-generation Fibonacci structure $(ABA)^{m}$ with the number of periods, $m$,
is shown in Fig. \ref{Fig.1}.
Generally, in the vicinity of the $K$ point and in the presence of a potential $V(x)$,
the charge carriers are described by the Dirac-like equation,
$
\hat{H}=-i\hbar v_{F}\mathbf{\vec{\sigma} \cdot \vec{\nabla} }+V(x)
$
where the Fermi velocity $v_{F}\approx 10^{6}$m/s, and $\mathbf{\vec{\sigma} }%
=(\sigma _{x},\sigma _{y})$ are the Pauli matrices. Due to the translation invariance in the $y$ direction,
the solution of above equation
for a given incident energy $E$ and potential barrier $V_j$ can be presented as $\tilde{\Psi} (x,y) = \Psi (x)  e^{i k_{y}y}$ with
%
$$
\label{barrier}
\Psi (x) = \left[
 a_j e^{i q_{j}x} \left(\begin{array}{c}
      1 \\
      \frac{q_j +ik_y}{k_j} \\
   \end{array}\right)
+
b_j e^{-iq_{j}x}
 \left(\begin{array}{c}
      1 \\
      \frac{-q_j +ik_y}{k_j} \\
   \end{array}\right)
 \right],
$$
where $k_j=(E-V_j)/\hbar v_F$, $k_{y}$ and $q_j$ are the $y$ and $x$ components of wavevector,
$q_{j}= \mbox{sign}(k_j) ({k^2_{j}-k_{y}^{2}})^{1/2}$ for $ k^2_{j}> k^2_{y}$,
otherwise $q_{j}=i  (k^2_{y}-k_{j}^{2})^{1/2}$, and $a_j$ ($b_j$) is the amplitude of the forward (backward)
propagating wave.
The wave functions at any two positions $x$ and $x +\Delta x $ inside the $j$th potential can
be related via the transfer matrix \cite{Wang2010}:
\begin{equation}
\label{barrier}
M_j =  \left(\begin{array}{cc}
      \frac{\cos(q_j \Delta x - \theta_j)}{\cos \theta_j} & i \frac{\sin(q_j \Delta x)}{\cos \theta_j}
      \\
      i \frac{\sin(q_j \Delta x)}{\cos \theta_j} & \frac{\cos(q_j \Delta x + \theta_j)}{\cos \theta_j}
   \end{array}\right),
\end{equation}
with $\theta_j = \arcsin(k_y/k_j)$. As a result, the transmission coefficient $t=t(E,k_y)$ is found to be
\beq
t =\frac{2 \cos \theta_0}{(m_{22} e^{- i\theta_0}+m_{11} e^{i\theta_t})-m_{12} e^{ i(\theta_t-\theta_0)}-m_{21}},
\eeq
where $\theta_0$ and $\theta_t$ are incidence and exit angles (see Fig. \ref{Fig.1}), and $m_{ij} (i,j=1,2)$ is
the matrix element of total transfer matrix, $X_N= \prod^{N}_{j=1} M_j$, connecting the incident and exit ends, and $N$ is the total number of
layers of the graphene superlattice.
Once the transmission coefficient is obtained,
the total conductance $G$ of the system at zero
temperature is given as follows,
$
G=G_{0}\int_{0}^{\pi/2} T \cos \theta_0
d \theta_0
$ \cite{Datta},
where $T=|t|^2$ and $G_0 = 2 e^2 m v_F L_y/\hbar^2$ and $L_y$ is the width of the graphene stripe in the $y$ direction.
Meanwhile, the Fano factor is given by
$
F= \int_{-\pi/2}^{\pi/2}T(1-T) \cos \theta_0
d \theta_0/ \int_{-\pi/2}^{\pi/2} T \cos \theta_0
d \theta_0
$ \cite{Beenakker-PRL}.

\begin{figure}[]
\begin{center}
\scalebox{0.5}[0.52]{\includegraphics{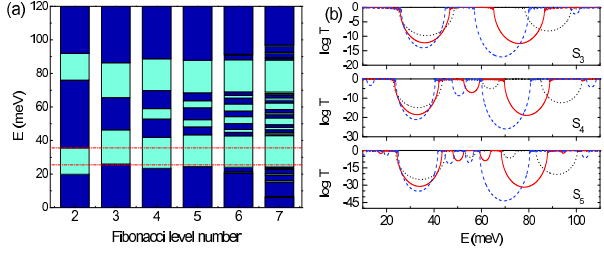}}
\caption{(Color online)  Energy band (a) and transmission spectrum (b) for the Fibonacci quasi-periodic GSLs
with $d_B/d_A=1$, where $V_A=50$ meV, $V_B=0$ meV, $\theta_0=20^{\circ}$, $m=16$,
(a): $d_A=20$ nm;
(b): $d_{A}=15$ nm (dotted black line), $d_{A}=20$ nm (solid red line), and $d_{A}=25$ nm (dashed blue line).
}
\label{energyband}
\end{center}
\end{figure}

Fig. \ref{energyband} shows the energy bands and transmission spectrum in various Fibonacci quasi-periodic GSLs (i.e., from $S_{2}$ to $S_{7}$).
Besides the distribution of energy bands like Cantor-like set \cite{Sena}, what we have discovered here is that
the zero-$\bar{k}$ gaps exist in all Fibonacci levels.
In Fig. \ref{energyband} (a),
there are several broad forbidden gaps opened for each Fibonacci level in the considered energy range. Among these forbidden gaps, we notice that
the position and size of zero-$\bar{k}$ gaps are almost robust against the Fibonacci levels.
In fact, the Fibonacci structure $S_2$ is exactly the GSL, $(AB)^m$.
The condition for zero-$\bar{k}$ gap is given by  $q_A d_A= - q_B d_B$ at $\theta_A=0$, 
which provides the DP, $E= V_A/(1+d_B/d_A)$, for the special case of $V_A \neq 0$ and $V_B =0$ \cite{Wang2010}.
For the higher Fibonacci level $S_{3}$ to $S_{7}$, the zero-$\bar{k}$ gaps become
stabilized with the fixed position and size, although the location of zero-$\bar{k}$ gaps is
slightly different from that for Fibonacci level $S_{2}$.
Furthermore, we demonstrate, in Fig. \ref{energyband} (b), that such
gap depends only on the ratio of lattice constants, and is insensitive to the lattice parameters.
On the contrary, the position and size of Bragg gaps in a higher energy range change sensitively with the Fibonacci
level and lattice parameters.

\begin{figure}[]
\begin{center}
\scalebox{0.40}[0.38]{\includegraphics{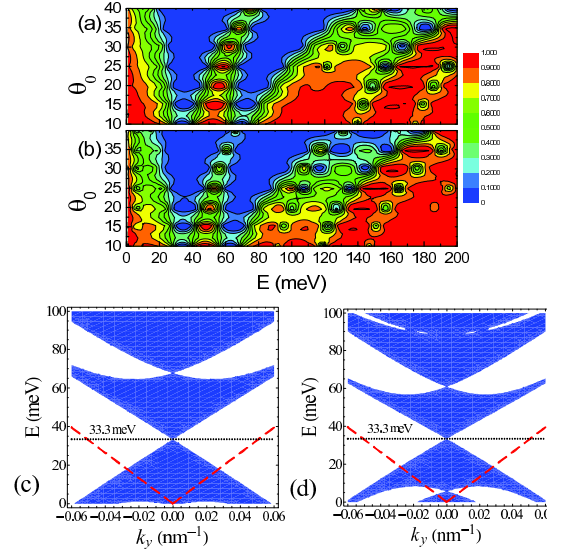}}
\\
\caption{(Color online) Transmission spectrum for the Fibonacci quasi-periodic GSL, $(ABA)^m$,
where $m=16$, (a) $d_{A}=20$ nm, (b) $d_{A}=25$ nm, and the other parameters are the same as those in Fig. \ref{energyband}.
(c) and (d) are the electronic band structures corresponding to (a) and (b).}
\label{transmission and bandstructure}
\end{center}
\end{figure}

Fig. \ref{transmission and bandstructure} (a) and (b) further display the influences of the incidence angle on the
transmission spectrum in the Fibonacci quasi-periodic GSL, $(AB)^m$, corresponding to Fibonacci sequence $S_3$.
It is apparent that the zero-$\bar{k}$ gap is independent of the lattice constants and is
weakly dependent on the incidence angle. To understand better,
the electronic dispersion at any incidence angle,
based on the Bloch's theorem, is written as,
\beqa
\label{dispersion}
\nonumber
\cos{(\beta_x \Lambda)} &\equiv& \frac{1}{2} \mbox{Tr} [M_A M_B M_A]= \cos{(2 q_{A}d_{A})}\cos{(q_{B}d_{B})}
\\ &+& \frac{\sin{\theta_{A}}\sin{\theta_B}-1}{\cos{\theta_A} \cos{\theta_B}} \sin{(2 q_{A}d_{A})}\sin{(q_{B}d_{B})},
\eeqa
where $\Lambda=2 d_A +d_B$ is the length of the unit cell. Therefore, the location of the DP
is given by $2 q_A d_A= - q_B d_B$ at $\theta_A=0$. For the structure considered here, ($V_A \neq 0$ and $V_B =0$),
the DP is exactly located at $E= V_A/(1+d_B/2d_A)$, which means that the zero-$\bar{k}$ gap depends only on the
ratio, $d_B/d_A$, instead of $d_A$ and $d_B$ themselves.
Fig. \ref{transmission and bandstructure} (c) and (d) show that
a band gap opens at $E = 33.3$ meV, which is different from $E=25$ meV for Fibonacci sequence $S_2$, in which $d_B/d_A=1$ and $V_A = 50$ meV.
In fact, DP for other Fibonacci sequences can be further calculated as $E=V_A/[1+ d_B/ \tau_{j} d_A]$,
where $\tau_{j}$ is the ratio of numbers of layer $A$ and $B$,
and $\lim_{j \rightarrow \infty} \tau_{j}= (1+\sqrt{5})/2=1.618$.
\begin{figure}[]
\begin{center}
\scalebox{0.6}[0.6]{\includegraphics{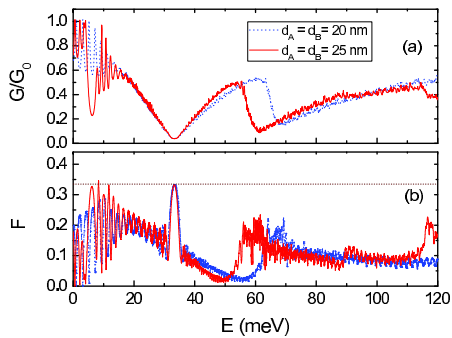}} \caption{(Color
online) Conductance (a) and Fano factor (b) as a function of Fermi
energy in Fibonacci quasi-periodic GSL, $(ABA)^m$,
where $m=16$, $d_A=20$ nm (dotted
blue line) and $d_B=25$ nm (solid red line), and the other
parameters are the same as those in Fig. \ref{energyband}.}
\label{conductance and fano}
\end{center}
\end{figure}

Fig. \ref{conductance and fano} shows the conductance and the Fano factor
with the difference lattice constants. Remarkably, the angular-averaged conductance
reaches the minimum value at the DP, while the Fano factor exists a
peak in the vicinity of DP with value approximately $F=1/3$ \cite{Beenakker-PRL,YuXian}.
The conductance and the Fano factor shows the
robust properties, since the DP does not shift, when the lattice constants $d_A$ and $d_B$ are
changed simultaneously.
\begin{figure}[ht]
\begin{center}
\scalebox{0.8}[0.8]{\includegraphics{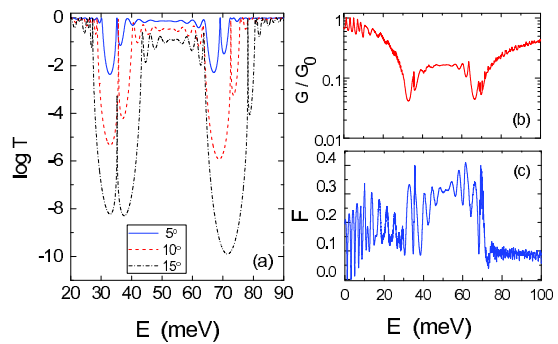}} \caption{(Color
online) Effect of defect mode on the transmission (a), conductance (b), Fano
factor (c) in GSL, where the defect $d_D=80$ nm, $V_{D}=55$
meV, $d_A=20$ nm, and the other parameters are the same as those in Fig.
\ref{energyband}.} \label{defect}
\end{center}
\end{figure}
In Fig. \ref{defect}, we further shed light on the effect of localized defect mode in GSL, $(ABA)^{8}D(ABA)^{8}$, where the defect layer $D$
with $d_D=80$ nm and $V_{D}=55$ meV. Compared to that inside the Bragg gap,
the defect mode inside the zero-$\bar{k}$ gap remains almost invariant with the incidence angles and lattice constants.
Due to the existence of the defect mode, the conductance is greatly enhanced, while
the Fano factor is strongly suppressed, as shown in Fig. \ref{defect}. This suggests
that the electron transport can be modulated by the defect mode.


In summary, using the transfer matrix method, we have investigated
the electronic band gap and transport in the Fibonacci quasi-periodic GSL.
It is shown that the zero-$k$ gap and the defect mode are robust against the lattice constants and incidence angle,
which is useful to control electron transport.
We hope such Fibonacci structure will
have applications in graphene-based electronic omnidirectional reflector and filters.


This work was supported by the NSFC (Grant Nos. 60806041 and 61176118),
and the Shanghai Leading Academic Discipline Program
(Grant No. S30105). X. C. also acknowledges Juan de la Cierva
Programme, the Basque Government (Grant No. IT472-10) and MICINN (Grant No. FIS2009-12773-C02-01).



\begin{thebibliography}{99}

\bibitem{Castro} A. H. Castro Neto, F. Guinea, N. M. R. Peres, K. S. Novoselov, A. K. Geim,
Rev. Mod. Phys. \textbf{81}, 109 (2009).


\bibitem{Novoselov} K. S. Novoselov, A. K. Geim, S. V. Morozov, D. Jiang, Y. Zhang, S. V. Dubonos, I. V. Grigorieva, A. A. Firsov,
Science \textbf{306}, 666 (2004).


\bibitem{Novoselov-GMJK} K. S. Novoselov, A. K. Geim, S. V. Morozov, D. Jiang, M. I. Katsnelson, I. V. Grigorieva, S. V. Dubonons, and A. A. Firsov,
Nature (London) \textbf{438}, 197 (2005).

\bibitem{Zhang-TS} Y. Zhang Y. W. Tan, H. L. Stromer, and P. Kim,
Nature (London) \textbf{438}, 201 (2005).

\bibitem{Gusynin-S} V. P. Gusynin, and S. G. Shararpov, Phys. Rev. Lett. \textbf{95}, 146801 (2005).

\bibitem{Katsnelson-NG} M. I. Katsnelson, K. S. Novoselov, and A. K. Geim, Nat. Phys. \textbf{2}, 620 (2006).

\bibitem{Young} A. F. Young and P. Kim, Annu. Rev. Condens. Matter Phys. \textbf{2}, 101 (2011).

\bibitem{Chen-APL}  X. Chen and J.-W. Tao, Appl. Phys. Lett. \textbf{94}, 262102 (2009).
\bibitem{Cheianov} V. V. Cheianov and V. I. Fal'ko, Phys. Rev. B \textbf{74}, 041403(R) (2006).





\bibitem{Meye} J. C. Meyer, C. O. Girit, M. F. Crommie, and A. Zettl,
Appl. Phys. Lett. \textbf{92}, 123110 (2008).

\bibitem{Marchini} S. Marchini, S. G\"{u}nther, and J. Wintterlin, Phys. Rev. B \textbf{76}, 075429 (2007).

\bibitem{Vazquez} A. L. Vazquez de Parga, F. Calleja, B. Borca, M. C. G. Passeggi, Jr., J. J. Hinarejos, F. Guinea, and R. Miranda,
Phys. Rev. Lett. \textbf{100}, 056807 (2008).



\bibitem{Bai} C.-X. Bai and X.-D. Zhang, Phys. Rev. B \textbf{76}, 075430 (2007).


\bibitem{Barbier2009} M. Barbier, F. M. Peeters, and P. Vasilopoulos,
Phys. Rev. B \textbf{80}, 205415 (2009); \textbf{81}, 075438 (2010).

\bibitem{Brey} L. Brey and H. A. Fertig, Phys. Rev. Lett. \textbf{103}, 046809 (2009).

\bibitem{Park2008} C. H. Park, L. Yang, Y. W. Son, M. L. Cohen, and S. G. Louie,
Phys. Rev. Lett. \textbf{101}, 126804 (2008).



\bibitem{Wang2010} L.-G. Wang and S.-Y. Zhu, Phys. Rev. B \textbf{81}, 205444 (2010); L.-G. Wang and X. Chen, J. Appl. Phys. \textbf{109}, 033710 (2011).

\bibitem{YuXian} X.-X. Guo, D. Liu, and Y.-X. Li, Appl. Phys. Lett. \textbf{98}, 242101 (2011).

\bibitem{Abedpour} N. Abedpour, A. Esmailpour, R. Asgari, and M. R. R. Tabar,
Phys. Rev. B \textbf{79}, 165412 (2009).

\bibitem{Bliokh} Y. P. Bliokh, V. Freilikher, S. Savel'ev, and F. Nori,
Phys. Rev. B \textbf{79}, 075123 (2009).


\bibitem{Mukhopadhyay} S. Mukhopadhyay, R. Biswas, and C. Sinha, Phys. Status. Solidi (b) \textbf{247}, 342 (2009).
\bibitem{Sena} S. H. R. Sena, J. M. Pereira Jr, G. A. Farias, M. S. Vasconcelos, and E. L. Albuquerque,
J. Phys.: Condens. Matter. \textbf{22}, 465305 (2010).


\bibitem{Li-PRL} J. Li, L. Zhou, C. T. Chan, and P. Sheng,
Phys. Rev. Lett. \textbf{90}, 083901 (2003).

\bibitem{Alfonoso} A. Bruno-Alfonoso, E. Reyes-G\'{o}mez, S. B. Cavalcanti, and L. E. Oliveira,
Phys. Rev. A \textbf{78}, 035801 (2008).
\bibitem{Zhang} L.-W. Zhang, K. Fang, G.-Q. Du, H.-T. Jiang, and J.-F. Zhao,
Opt. Commun. \textbf{284}, 703 (2011).


\bibitem{Datta} S. Datta, \textit{Electronic Transport in Mesoscopic Systems} (Cambridge University Press, Cambridge, England, 1995).

\bibitem{Beenakker-PRL} J. Tworzyd{\l}o, B. Trauzettel. M. Titov, A. Rycerz, and C. W. J. Beenakker,
Phys. Rev. Lett. \textbf{96}, 246802 (2006).


\end{thebibliography}
\end{document}